# Experimental investigation of the "ratchet" effect in a two-dimensional electron system with broken spatial inversion symmetry

S. Sassine,[1,2,*] Yu. Krupko,[1] J.-C. Portal,[1,2,4] Z. D. Kvon,[3] R. Murali,[5] K. P. Martin,[5] G. Hill,[6] and A. D. Wieck[7]

[1] GHMFL, CNRS, BP 166, F-38042, Grenoble Cedex 9, France
[2] INSA, 135 Avenue de Rangueil, F-31077, Toulouse Cedex 4, France
[3] Institute of Semiconductor Physics, Novosibirsk 630090, Russia
[4] Institut Universitaire de France, 103 Bd. St. Michel, F-75005, Paris, France
[5] Microelectronics Research Center–Georgia Institute of Technology, 791 Atlantic Dr NW, Atlanta GA30332, USA
[6] Department of Electronic and Electrical Engineering–University of Sheffield, S13JD Sheffield, UK
[7] Lehrstuhl fur Angewandte Festkörperphysik, Ruhr-Universität Bochum D-44780, Bochum, Germany

(Dated 28 February 2008)

We report on experimental evidence of directed electron transport, induced by external linear-polarized microwave irradiation, in a two-dimensional spatially-periodic asymmetrical system called "ratchet". The broken spatial symmetry was introduced in a high mobility two-dimensional electron gas based on AlGaAs/GaAs heterojunction, by patterning an array of artificial semi-discs-shaped antidots. We show that the direction of the transport is efficiently changed by microwave polarization. The dependence of the effect on magnetic field and temperature is investigated. This represents a significant step towards the realization of new microwave detectors and current generators.

# I. INTRODUCTION

In the absence of any static forces, the appearance of a directed transport induced by external energy sources in asymmetric systems is known as the "ratchet" effect. This effect has a generic nature and it has been observed in various physical systems.[1–3] It also has important applications in biological systems.[4,5] In condensed matter physics, the directed transport in externally driven systems with naturally broken spatial inversion symmetry has been studied extensively for many years. A number of phenomena including photogalvanic effect,[6,7] surface photovoltaic effect,[8] mesoscopic photovoltaic effect,[9,10] spin-galvanic effect[11] and many others[12] were investigated. As for artificially broken symmetry and related phenomena in semiconductor-based devices, several studies have been reported[13–16] (in which the symmetry was mostly broken by depletion gates-defined single open quantum dot). Experimental studies on tunneling ratchets were performed on semiconductor heterostructures by confining electrons to an asymmetric conducting channel and under applied source-drain bias voltage.[17] Also, previous experimental studies on "ratchet" electron transport in semiconductor heterostructures with broken symmetry have been reported[18,19] demonstrating the existence of the effect, but no detailed investigations have been done so far. In these studies,[18,19] lateral photo-voltages were obtained in arrays of triangular antidots on semiconductor heterostructures under far-infrared irradiation[18] (up to 0.2 µV) at a temperature of 4 K, and under microwave irradiation[19] (up to 50 µV) at room temperature. Note also that directed current was recently observed in asymmetrical open ballistic dot under microwave radiation.[20]

In this study, we used asymmetric antidot arrays (having several thousands of antidots) fabricated on a high mobility two-dimensional electron gas (2DEG) based on a semiconductor heterojunction. These antidot lattices are well known examples where the scattering of charged particles (electrons or holes) is controlled by the artificial scattering potential (antidots). Symmetry breaking is obtained by choosing a specific (non-circular) shape for the antidots, for example a semi-disc. We demonstrate experimentally that such systems show large directed electron motion under irradiation with linear-polarized microwaves (MW), which is related to the fact that the total scattering of electrons in the 2DEG on the asymmetric antidots leads to a directed flow. The crucial questions about the dependence of the direction of the induced current on the orientation of the linear-polarization, and the non-existence of the effect in lattices with symmetric scatterers, are addressed here for the first time. Also, the dependence of this "ratchet" electron transport on some important parameters such as magnetic field and temperature has been established. The aim of our study is to perform experimental investigations of the electronic "ratchet" effect in detail, to compare our experimental results to recent theoretical works[21–24] (which support our experiments) and to exploit semiconductor engineering to establish the sensitivity of the system to MW polarization.

## II. EXPERIMENTAL

Following established theoretical ideas,[21–24] we fabricated a periodic array of asymmetrical semidisc-shaped antidots (positioned in a hexagonal lattice) on a high mobility 2DEG (see inset to Fig. 1). In order to show the key role of the system asymmetry in the origin of the "ratchet" effect, lattices with the same parameters but with symmetrical circular antidots were fabricated on the same sample using the same technological process. The lithographic parameters of the lattices were: the period $a = 1.5$ μm and the antidot radius $r = 0.5$ μm.

Our samples have been fabricated on the basis of a molecular-beam-epitaxy-(MBE)-grown AlGaAs/GaAs modulation-doped heterojunction, having 2DEG at a depth of 117 nm below the surface. Three consecutive Hall bars with lateral sizes $250 \times 50$ μm$^2$ were fabricated in series on top of every sample, using conventional photolithography techniques. The array of semicircular antidots has been fabricated in one of these three Hall bars and that of circular antidots in a second one, by means of electron beam lithography and subsequent plasma etching. The third Hall bar on every sample was left unaltered (unpatterned, i.e. without antidots) in order to perform Hall measurements of the unperturbed 2DEG. The fabrication process does not alter the electron density since Shubnikov-de-Haas oscillations are measured to be the same (same positions of minima) in the patterned and the unpatterned part (see Fig. 1).

Magneto-transport and photo-voltage measurements have been performed while implementing the study: the magneto-transport was measured by a standard low-frequency ac technique with low current (13 Hz, 0.1 μA) to identify the 2DEG parameters and to study the transport in the antidot lattice; the dc photo-voltage measurements were carried out with a high accuracy digital multimeter to measure the signal directly from the "ratchet" lattice (see left inset of Fig. 2 where the voltage signal was measured using both the contact pair 1,2 and the pair 3,4). The measurements have been performed in low magnetic fields (normal to the 2DEG plane) at temperatures 1.4–100 K, using linear-polarized MW irradiation from a "carcinotron" generator tunable in the 33–50 GHz frequency range.

To provide minimal damping of the MW power, a circular-section brass tube has been used to guide MW to the sample. In order to concentrate the MW power on the sample area and, at the same time, to control the linear polarization of the MW irradiation, special brass insets have been fabricated. They reduce the transmission-line internal profile from the circular to a rectangular-waveguide and vice versa. Placing the rectangular waveguide profiles on both sides of the transmission line, linear polarization of the microwaves is obtained. Finally, the sample was placed at a distance of 1–2 mm in front of the waveguide output in a Variable Temperature Cryostat. In order to exclude the influence of the MW irradiation on the sample contacts, a metallic diaphragm was placed between the sample and the waveguide as close as possible to the sample. The diaphragm's hole was situated precisely above the antidot lattice.

## III. RESULTS AND DISCUSSION

The starting 2DEG has the electron density $N_s = (2-3) \times 10^{15}$ m$^{-2}$ and mobility $\mu = (2-3) \times 10^2$ m$^2$/Vs at 1.5 K. This corresponds to an electron mean free path of (15–30) μm which is much larger than the antidot spacing. The transport properties should therefore be dominated by the antidots scattering potential at low temperatures. Indeed, Fig. 1 shows that the lattice resistivity at zero magnetic field is 480 Ω/square (blue curve), about 50 times larger than that of the unpatterned 2DEG (red curve). Four well developed commensurability peaks are also present in the magneto-resistivity of this lattice. They relate to the formation of localized trajectories around either a single antidot or a group of antidots for specific values of the magnetic field.[25–27] The most pronounced peak (number 3 in Fig. 1) corresponds to the condition $2R_c = a$ ($R_c$ is the cyclotron radius) and correlates to the formation of trajectories surrounding a single antidot. The other peaks correspond to trajectories formed in-between three antidots (peak number 4) and the splitting of one commensurability peak corresponding to the trajectory surrounding 7 antidots[25–27] (1 and 2). Figure 1 therefore demonstrates experimentally that the crucial condition for rectifying transport in the "ratchet" lattice is that antidots play a dominant role as asymmetric intentionally ordered artificial scatterers at low temperatures. Electrons can be considered to move ballistically between collisions with antidots. In the following we will discuss the "ratchet" effect in this potential.

According to the second law of thermodynamics there is no directed transport in spatially periodic asymmetric systems at thermal equilibrium.[28] The introduction of the linear-polarized MW may drive the system ("ratchet" lattice with 2DEG) out of equilibrium and create a directed electron flow whose direction is related to the system configuration. This unusual phenomenon is based on the interplay of space asymmetry and external periodic driving force. We claim that our system is in thermal equilibrium at 1.5 K. According to theory,[22,23] a strong "ratchet" effect appears (with no limit on MW frequency) when the energy of the MW photon is larger than the energy spacing $\Delta$ inside one antidot lattice cell: $\hbar\omega > \Delta \approx 2\pi\hbar^2/(ma^2)$, where $\omega$ is the MW frequency, $\hbar$ the Planck's constant and $m$ the electron effective mass. In the absence of any external current or bias applied to the sample, the appearance of a dc-voltage of a few mV induced by the linear-polarized MW irradiation was observed in the "ratchet" antidot lattice. In Fig. 2, the key results of the rectification experiments are presented: the magnetic field dependence of the dc-voltage measured in the "ratchet" antidot lattice under MW irradiation of frequency 42.7 GHz. The red curve corresponds to the case when the vector of linear polarization (driving electric field $\vec{E}$) was oriented along the x-axis (see left inset of Fig. 2) and the blue curve corresponds to the case when $\vec{E}$ was oriented along the y-axis. Importantly, at zero magnetic field, the sign of the signal is opposite for the two orientations of polarization. When $\vec{E}$ coincides with the y-axis, it forces electrons to oscillate vertically and scatter in majority on the semicircular side of the antidots (see left inset of Fig. 2). This leads electrons to move to the right thereby causing a negative sign of the signal ($U_{ph} = U_A - U_B$). On the contrary, when $\vec{E}$ is oriented along the x-axis, it pushes electrons to oscillate horizontally and the total scattering in this case is dominated by the plane side of antidots and electrons flow to the left (in this case, scattering on the semicircular side doesn't give any directed flow). In other words, the signal changes sign! This demonstrates that it is possible to control the direction of the transport using the direction of MW polarization. For MW polarized along the x-axis, the average transport goes along the –x direction, while for MW polarized along the y-axis the direction of transport is inversed. In both cases the current flows along the x direction which is the direction where spatial inversion symmetry of the system is broken. Importantly, the "ratchet" signal is absent in a lattice of symmetrical circular antidots (orange and violet curves in Fig. 2). This further confirms that broken spatial symmetry is a necessary ingredient to the "ratchet" effect. These are important results of our study and they are in good qualitative agreement with theoretical predictions.[21–24] However, the value of the signal at zero magnetic field for $\vec{E}$ along the x-axis is not equal to that for $\vec{E}$ along the y-axis. The possible reason for this is most probably that MW irradiation can excite plasmon resonance[29,30] which may affect the measured signal somehow. Therefore, further investigations (for different Hall bar geometries) of this problem are necessary.

The power dependence of the dc signal (at zero magnetic field) is shown in the right inset of Fig. 2 for both directions of polarization. This dependence is quasi-linear. This is related to the fact that the electron flow velocity, and thus the strength of the current, increases approximately linearly with the increase of MW power. Such behaviour is expected by theory.[23] Note that the output power of the MW generator is used as x-axis in this graph. Because of the power losses in the waveguide and at the sample surface, only a small portion of the power from the MW source actually couples into the patterned 2DEG. This power is estimated to be about 20 µW for the curves presented in Fig. 2. Thus, the MW electric field ($E_\omega$) in the "ratchet" lattice is about 9 V/cm. Let us now compare the experimental value of the directed current to theoretical calculations. The theory[23] gives the following simple expression for the rectification current at $\omega\tau < 1$ ($\tau$ is the transport time in the lattice, in our case $\omega\tau \approx 0.5$):

$$I_{rect} \approx 0.13 e N_s V_F (erE_\omega/E_F)^2 \times D \qquad (1)$$

where $e$ is the electron charge, $V_F$ the Fermi velocity and $D$ is the width of the lattice (50 µm). We obtain $I_{rect} = 0.7$ µA using Eq. (1) which is comparable to the experimental value ($I_{rect} = U_{ph}/R_{lat}$ where $R_{lat}$ is the resistance of the lattice) of about 0.9 µA (when $\vec{E}$ is along the x-axis) and not so far from that of about 8 µA (when $\vec{E}$ is along the y-axis).

Let us now discuss the magnetic field dependence of the dc signal. At low magnetic fields (- 0.2 T < $B$ < 0.2 T), the signal for both directions of polarization strongly decreases. This is because as the magnetic field increases, classical cyclotron orbits decrease in size and the interaction of the electrons with the antidots decreases. Also, we observe the presence of some remnants of commensurability peaks in this interval of magnetic field. For example, the magnetic field position of the first peak of the blue curve (to the left or to the right) matches with that of the fundamental commensurability peak (peak number 3 in Fig. 1). At higher magnetic fields ($B > |0.2|$ T), the effect vanishes for any direction of MW polarization since the magnetic field beyond the quantum limit causes breaking of the linear-trajectories geometry.[25] In another picture, the classical cyclotron orbits are then smaller than the antidot spacing ($2R_c < a$) and the scattering on antidots is totally suppressed. The value of the magnetic field corresponding to the suppression of the effect $B_s \approx 0.2$ T, agrees well with theoretical calculations.[23] Also, it is worth noting the nonzero signal for $\vec{E}$ along the y-axis in one direction of the magnetic field (right part of the blue curve). Probably, this is a signature of a new kind of photo-magnetic effect. It will be discussed in more details elsewhere.

Finally, Fig. 3 presents the temperature dependence of the dc signal (at zero magnetic field), plotted together with the mobility of the "ratchet" lattice and the unpatterned 2DEG. As expected, the mobility of the 2DEG has a behaviour comparable to high mobility samples[31] (i.e. phonon scattering is dominant down to a few Kelvin). On the contrary, the mobility in the patterned part is two orders of magnitude smaller and almost constant below 70 K. This means that scattering on the antidots entirely controls the transport for $T < 70$ K. At higher temperatures, isotropic phonon scattering becomes dominant, as can be seen in the rectification signal that vanishes at $T \approx 70$ K. In fact, the electron mean free path is 19 μm at 1.5 K and it decreases with increasing temperature. At 70 K, it becomes equal to 1.7 μm which is comparable to the period of our "ratchet" lattice, so electrons don't feel the "ratchet" lattice anymore. The correlation between the two signals is a strong indication of the "ratchet" effect which requires the mean free path to be larger than the antidot spacing. This too is in accordance with theory.[24]

## IV. CONCLUSION

In conclusion and based on our observation of the "ratchet rectification" phenomenon in lattices with relatively large mesoscopic period (1.5 µm) at liquid nitrogen temperatures, we believe that the "ratchet" effect has classical grounds and it is possible to observe it (under polarized MWs) at higher temperatures. To demonstrate this, it is necessary to fabricate a "ratchet" lattice with a smaller period (one order of magnitude smaller) that nowadays is not beyond the technical possibilities, on a 2DEG with electron mean free path larger than the lattice period at the desired temperature. Moreover, experiments at higher frequencies will be necessary. In turn, this effect offers interesting possibilities for making new electromagnetic radiation detectors sensitive to polarization operating in MW and terahertz frequency ranges. Also, the perspective to fabricate new micro-scale-sized current generators is proposed since our "ratchet" lattice acts as a current generator. Finally, artificial asymmetric antidot lattices can be considered as a prototype for transport in asymmetric molecular systems.

## ACKNOWLEDGEMENTS

The authors express their deep acknowledgements to A. D. Chepelianskii, D. L. Shepelyansky, M. V. Entin and V. T. Renard for very fruitful discussions. Also authors are very grateful to J. Florentin, A. Richard, H. Aubert and H. Granier for their technical support. The samples have thankfully been MBE-grown with the help of Dr. D. Reuter. This work was financially supported by ANR/PNANO-MICONANO and by CNRS/RAS (19046): PICS project (3862).

**FIGURE CAPTIONS**

FIG. 1. Magneto-resistivity traces of the antidot lattice (blue) and intact 2DEG part (red). An atomic force microscope image of the lattice including the etched profile (with sketch of the commensurability orbits 3 and 4) as well as the electron system parameters are shown in the two insets.

FIG. 2. Magnetic field dependence of the dc-voltage ($U_A - U_B$) measured in "ratchet" antidot lattice under MW irradiation having the vector of linear polarization parallel to the x-axis (red) and perpendicular to it (blue). Importantly, the sign of the signal is opposite for the two orientations of polarization at zero magnetic field. At higher magnetic fields ($B > 0.2$ T), the effect vanishes for any direction of MW polarization. The absence of the "ratchet" signal in a lattice of symmetrical circular scatterers (orange and violet) proves that broken symmetry is a necessary ingredient to the "ratchet" effect. The insets show the power dependence of the polarization-dependent signal (to the right) and the measurement configuration with a schematic representation of the scattering process (on one antidot) for both directions of polarization (to the left).

FIG. 3. Temperature dependence of the "ratchet" signal (at zero magnetic field) plotted together with temperature dependencies of the mobility in the sample for "ratchet" and unpatterned 2DEG (temperature and mobility axes are presented in logarithmical scales).

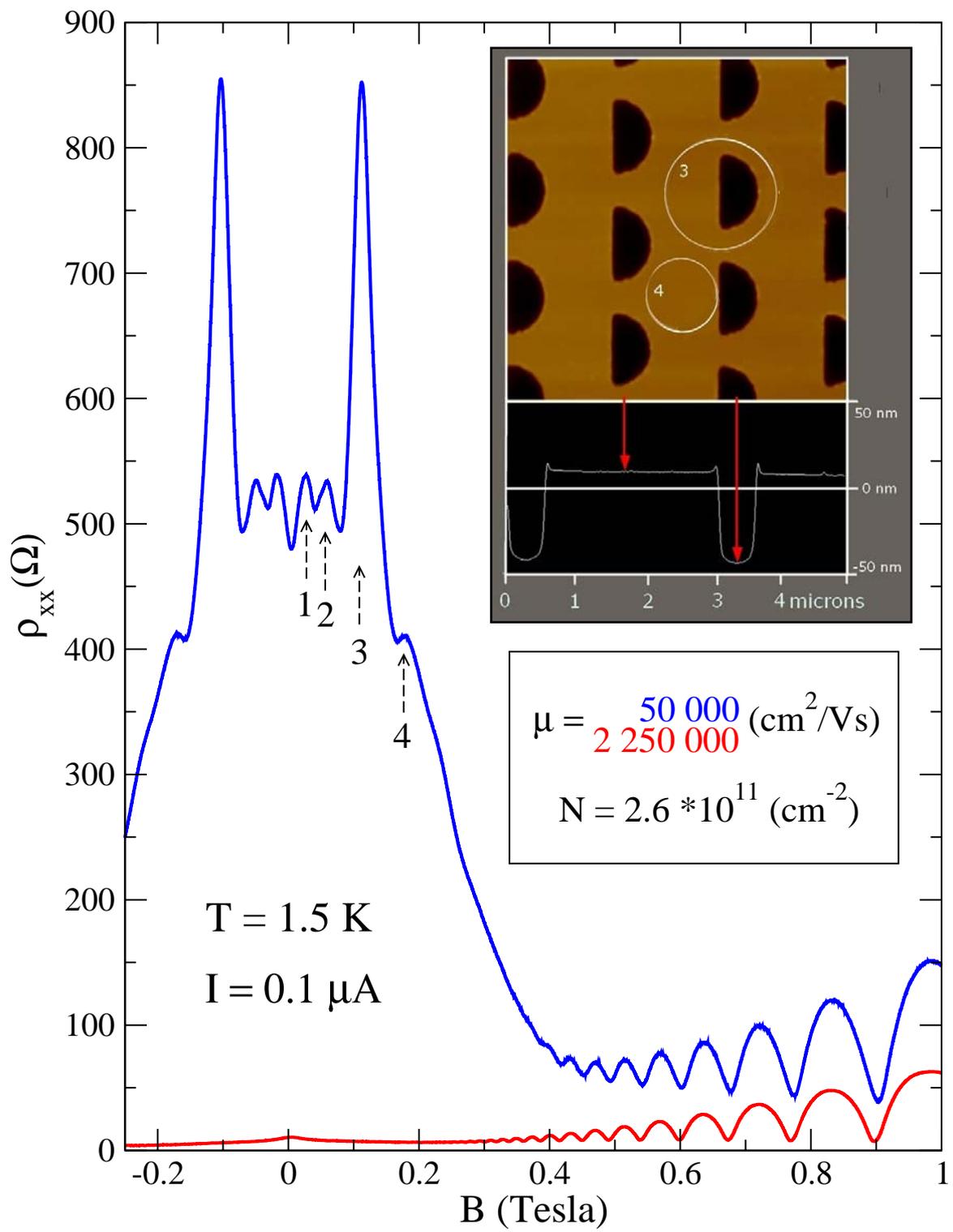

**FIG. 1.**

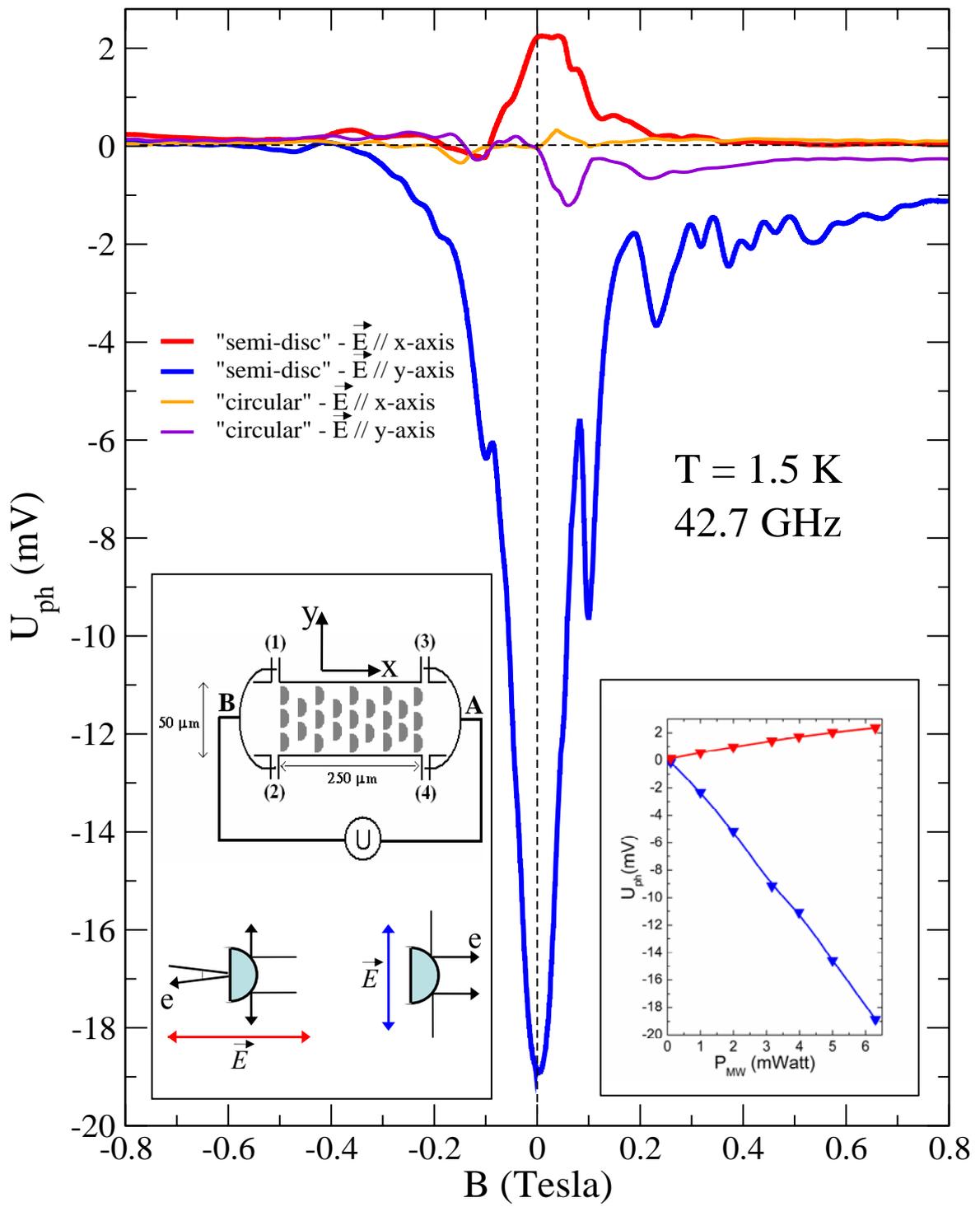

**FIG. 2.**

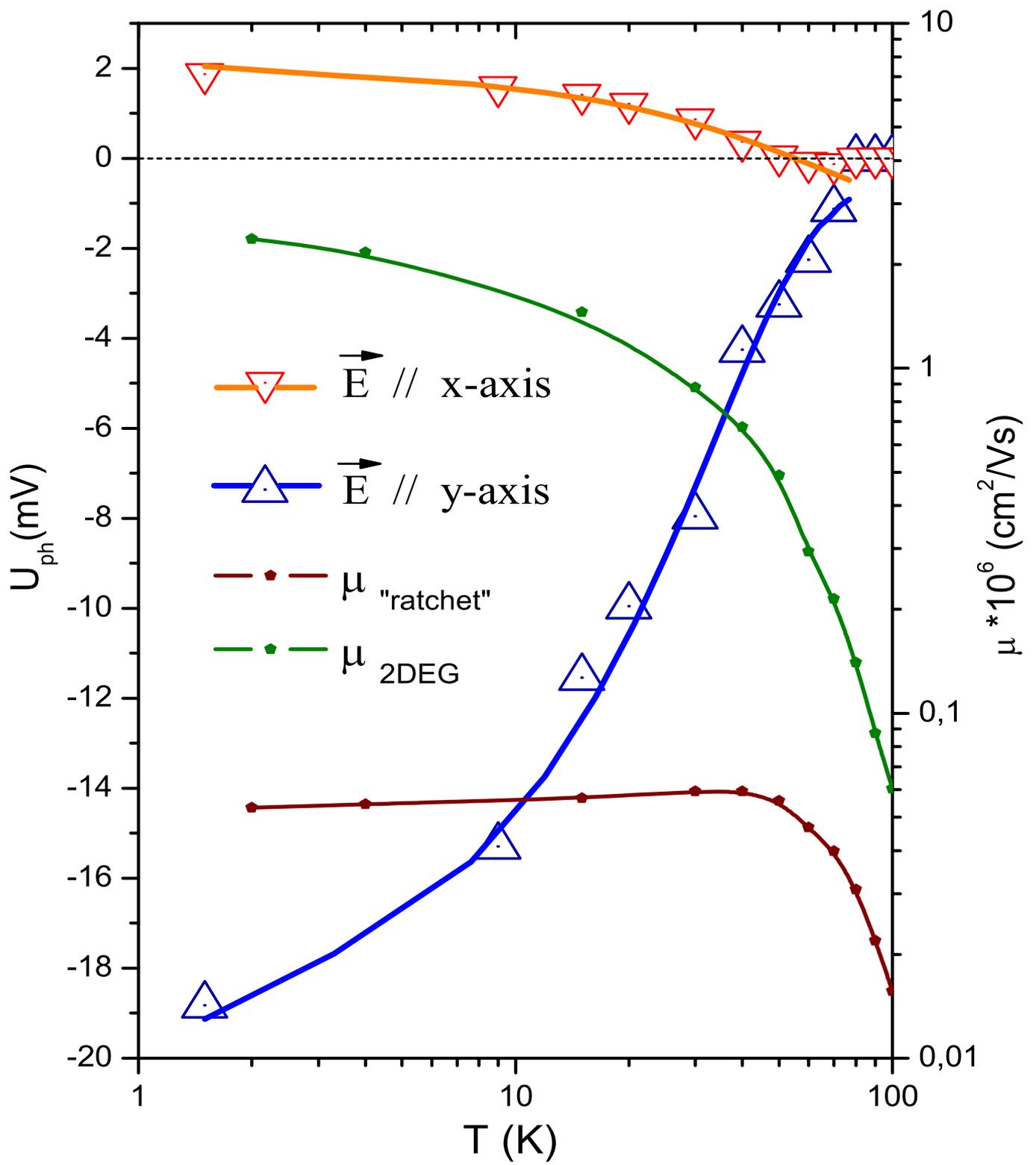

**FIG. 3.**